\input harvmac.tex
\vskip 1in
\Title{\vbox{\baselineskip8pt
\hbox to \hsize{\hfill}
\hbox to \hsize{\hfill IC/2000/58}}}
{\vbox{\centerline{ AdS/CFT Correspondence, Critical Strings}
\vskip 0.1in
{\vbox{\centerline{and Stochastic Quantization}}}}}
\centerline{Dimitri Polyakov\footnote{$^\dagger$}
{polyakov@ictp.trieste.it}}
\medskip
\centerline{\it The Abdus Salam International Centre for Theoretical Physics}
\centerline{\it Strada Costiera 11 }
\centerline{\it I-34014  Trieste, ITALIA}
\vskip .1in
\centerline {\bf Abstract}
In our previous paper we have shown that the NSR string sigma-model
with the massless  5-form vertex operator in $D=10$ NSR string theory:
$V_5{\sim}e^{-3\phi}\psi_0\psi_1\psi_2\psi_3\psi_t\bar
\partial{X^t}e^{ik^{||}x^{||}}$
($t=4,..9$)
reproduces the  correlators of the $N=4$ $D=4$ super Yang-Mills theory.
In particular, this implies that the sigma-model
with the $V_5$ operator in flat space-time should be the NSR analogue of
the GS string theory on $AdS_5\times{S^5}$.
This means that the $V_5$-operator plays the role of cosmological
constant, curving flat ten-dimensional space-time into 
that of $AdS_5\times{S^5}$
In the present paper we show  that dilaton beta-function
equation in such a sigma-model has the form of stochastic
Langevin equation with the non-Markovian noise.
The  worldsheet cutoff is identified with stochastic time and
the $V_5$-operator plays the role of the noise.
We derive the Fokker-Planck equation associated with
this stochastic process and show that the Hamiltonian
of the $AdS_5$ supergravity defines the distribution satisfying
this Fokker-Planck equation.
This means that the  dynamical compactification
of the space-time on $AdS_5\times{S^5}$ occurs as a result
of the non-Markovian stochastic process, generated by the $V_5$-operator
noise.
This provides us with an insight into relations  
between holography principle and the concept of stochastic quantization
from the point of view of critical string theory.
\Date{May 2000 {\bf PACS: $04.50.+h$;$11.25.Mj$.}}
\vfill\eject
\lref\verlinde{E.Verlinde, H.Verlinde, hep-th/9912018}
\lref\verlind{J. de Boer, E.Verlinde, H.Verlinde, hep-th/9912018}
\lref\myself{D.Polyakov, {\bf Phys.Lett. B 469 (1999) 103}}
\lref\periwal{L.Lifschyts, V.Periwal, hep-th/0003179, JHEP 0004: 026, 2000}
\lref\hirano{S.Hirano, hep-th/9910256}
\lref\ampr{A.M.Polyakov, V.Rychkov, PUPT-1917, hep-th/0002106}
\lref\huffel{P.Damgaard, H.Huffel, {\bf Phys.Rep.152 (1987) 227}}
\lref\hhuffel{P.Damgaard, H.Huffel, Eds., Stochastic Quantization, 
World Scientific (1988)}
\lref\ampf{S.Gubser,I.Klebanov, A.M.Polyakov, 
{\bf Phys.Lett.B428:105-114}}
\lref\amps{A.M.Polyakov,hep-th/9809057}
\lref\myselff{D. Polyakov, hep-th/9812044, {\bf Phys.Lett.B470:103 (1999)}}
\lref\parisi{G.Parisi, Y.S.Wu, {\bf Sci. Sinica 24 (1981) 484}}
\lref\malda{J.Maldacena, Adv.Theor.Math.Phys.2 (1998)
231-252, hep-th/9711200}
\lref\maldac{J.Malfacena, hep-th/9803002, {\bf Phys.Rev.Lett. 80: 4859
(1998)}}
\lref\wit{E.Witten{\bf Adv.Theor.Math.Phys.2:253-291,1998}}
\lref\myselfin{D.Polyakov, in progress}
\lref\dhoker{E.D'Hoker, D.Freedman, S.Mathur, A.Matusis, 
{\bf Nucl.Phys.B 456:96 (1999)}}
\lref\sonn{C.Sonnenschein, hep-th/0003032}
\lref\colomb{J.-F. Colombeau ``Multiplication of Distributions...''
Springer-Verlag, 1992}
\lref\van{C. Van De Bruck, gr-qc/0001048}
\centerline{\bf Introduction}
The holography principle ~\refs{\malda, \amps,\ampf, \wit}
relates degrees of freedom of quantum field
theories in D and $D+1$ dimensions. The remarkable example  is the
AdS/CFT correspondence which relates the AdS supergravity to
supersymmetric conformal field theories  on the AdS boundary.
In particular, in case of $D=4$ this implies the correspondence  between
five-dimensional AdS supergravity and $N=4$ super Yang-Mills theory.
Another, seemingly unrelated example  connecting 
 field theories in $D$ and $D+1$ dimensions 
is known from the concept of stochastic quantization.
In the approach of stochastic quantization ~\refs{\parisi}
(for comprehensive review see also ~\refs{\huffel, \hhuffel})
 the dynamics of a given
D-dimensional quantum field theory is studied  by introducing
the white noise to the field theory, or coupling it to
infinite thermal reservoir. The interaction with the reservoir 
breaks the thermal equilibrium;
as a result the field theory starts behaving like a brownian-like
object,  drifting in an $additional$ ``fictitious'' dimension, known
as stochastic time $\tau$. In the limit of infinite stochastic time
this non-equilibrium $D+1$-dimensional system with certain
 Fokker-Planck distribution returns to the initial equilibrium state;
the
logarithm of the equilibrium limit of the FP distribution is
identified with the  action of the original 
$D$-dimensional field theory.
The dynamics of this $D+1$-dimensional stochastic field theory
is described by the correlators
$<\phi(x_1,\tau_1)....\phi(x_n,\tau_n)>$ where $\phi(x,\tau)$ are
the solutions of the Langevin equation:

\eqn\lowen{{{d\phi}\over{d\tau}}=-{{\delta{S_E}}\over{\delta{\phi}}}
+\eta(x,\tau)}
 Here $S_E$ is the action of the equilibrium
$D$-dimensional theory and $\eta$ is the white noise.
To obtain correlation functions 
the product of these solutions (as functions of $\eta$,
obtained by iterations)
must be averaged over the white noise with the gaussian distribution.
This gives rise to the $D+1$-dimensional stochastic diagrams which, 
in the limit
of $\tau\rightarrow\infty$ reproduce the Feynman graphs of the 
D-dimensional theory. 
 
In the present paper we will try to show that 
 these two approaches , the stochastic quantization and the 
holography principle, are in fact related to each other 
in a rather intriguing way and that such a
relation may be observed and explored
 from string theory point of view.
 The key element of our construction is so called brane-like
sigma-model
 which may be understood as the NSR analogue
of the GS superstring theory on $AdS_5\times{S^5}$,
as the analysis of its correlation functions shows ~\refs{\myself}.
 The action for this model is given by:
\eqn\grav{\eqalign{{S_{b.l.}}=\int{d^2}z
\partial{X^\mu}\bar\partial{X_\mu}+\psi^\mu\bar\partial\psi_\mu
+\bar\psi^\mu\partial{\bar\psi_\mu}\cr+\lambda_0\epsilon_{a_1a_2a_3a_4}
\int{{d^4{k^{||}}}\over{{k^{||}}^4}}
e^{-3\phi}\psi^{a_1}\psi^{a_2}\psi^{a_3}\psi^{a_4}\psi_t\bar\partial{X^t}
e^{ik^{||}X}+c.c.+ghosts}}
Here $X^\mu(z,\bar{z})(\mu=0,...,9)$ are space-time coordinates,
$\psi_\mu$ and $\bar\psi_\mu$ are NSR worldsheet fermions
and $\phi$ is bosonized superconformal ghost field.
Furthermore, the space-time index $\mu$ is split in the $4+6$ way,
so that $a=0,...,3$ and $t=4,...,9$.
In other words, the sigma-model (2) is given by NSR string theory
action in  flat space-time with the potential term related to 
 the closed-string brane-like vertex operator
\eqn\lowen{V_5(z,\bar{z},k^{||})={{\lambda_0}\over{{{k^{||}}^4}}}
\epsilon_{a_1...a_4}e^{-3\phi}\psi_{a_1}...\psi_{a_4}\psi_t\bar\partial
{X^t}e^{ik^{||}X}(z,\bar{z})}
which we will also refer to as $V_5$-operator in the rest of the paper.
The properties of the sigma-model with the $V_5$-operator have been studied
in ~\refs{\myself} where the relevance of this vertex operator
to non-perturbative $D3$-brane  dynamics has been shown.
It is important that BRST invariance condition for the $V_5$-operator
(3) requires that its propagation is confined to four dimensions.
Namely, to insure the BRST invariance,
the momentum $k^{||}$ must be polarized along the  $a_1,...a_4$ directions.
Moreover, the condition of the worldsheet  conformal invariance
(preserving the conformally invariant form of the O.P.E. between
two stress-energy tensors corresponding to the action (2) or
the vanishing of the beta-function in the lowest order of string
perturbation theory) requires that the space-time scalar field
$\lambda(k^{||})$, corresponding to the $V_5$-operator, should behave as
\eqn\lowen{\lambda(k^{||})\sim{{\lambda_0}\over{{{k^{||}}^4}}}}
where $\lambda_0$ is constant.
As has been argued in ~\refs{\myself} the role of the $V_5$ operator
is that it transforms the flat ten-dimensional space-time vacuum
into that of $AdS_5\times{S^5}$, thus connecting two maximally supersymmetric
backgrounds in ten dimensions. This is because  adding the $V_5$-term to
the sigma-model action (2) is in fact equivalent to introducing the $D3$-brane
in the theory. As a result, one may explore string
theory in the AdS background (and  consequently,
 the large N limit of gauge theory) by means of the brane-like
sigma-model (2) which technically lives in $flat$ ten-dimensional 
space-time. 
Using the correspondence  between local gauge invariant
operators in the large N Yang-Mills theory and 
massless vertex operators in string theory one may obtain
the large N correlators in gauge theory by computing  
scattering amplitudes of the BRST invariant vertices in the 
sigma-model (2).  For example, as the dilaton vertex operator $V_\varphi$
corresponds to the $Tr{F^2}$ field in gauge theory, the generating
functional for various correlation functions of the $Tr{F^2}$ 
operators is given by:
\eqn\grav{\eqalign{Z(\lambda_0,\varphi)=
\int{D}\lbrack{X}\rbrack{D}\lbrack\psi\rbrack{D}\lbrack{ghosts}\rbrack
f(\Gamma,N){exp}\lbrace\int{d^2}z
\partial{X^\mu}\bar\partial{X_\mu}+\psi^\mu\bar\partial\psi_\mu
+\bar\psi^\mu\partial\bar\psi_\mu\cr+\lambda_0\epsilon_{a_1a_2a_3a_4}
\int{{d^4{k^{||}}}\over{{k^{||}}^4}}
e^{-3\phi}\psi^{a_1}\psi^{a_2}\psi^{a_3}\psi^{a_4}\psi_t\bar\partial{X^t}
e^{ik^{||}X}+\int{d^{10}p}V_\varphi(p,z,\bar{z})\varphi(p)\rbrace
\cr+c.c.+ghosts}}
where $\varphi(p)$ is ten-dimensional space-time dilaton field.
The ``measure function'' $f(\Gamma,N)\sim{(1+N^2\Gamma^4)^{-1}}
\times{c.c.}$
( $\Gamma$ is picture-changing operator and $N$ corresponds
to the gauge group parameter) needs to be introduced
to the measure of integration to insure correct ghost number balance
on the sphere
and normalization of scattering amplitudes.
The two-point dilaton correlation function, corresponding to
the  generating functional (5) is given by
\eqn\grav{\eqalign{<V_\varphi(p_1)V_\varphi(p_2)>_{\sigma-model}
={{\delta^2{Z(\lambda_0,\varphi)}}\over{\delta\varphi(p_1)\delta\varphi(p_2)}}
|_{\varphi=0}}}
To compute this correlator we have to expand the functional (5)
in $\lambda_0$. The first non-trivial contribution has the order
of $\lambda_0^2$ and it is given by
\eqn\grav{\eqalign{A_{\lambda_0^2}(p_1,p_2)\sim
\lambda_0^2\int{{d^4{k_1^{||}}}\over{{k_1^{||}}^4}}
\int{{d^4{k_2^{||}}}\over{{k_2^{||}}^4}}<V_\varphi(p_1)V_\varphi(p_2)
V_5(k_1^{||})V_5(k_2^{||})>}}
where the four-point amplitude should be computed in the usual
NSR string theory in flat space-time.
In other words, this is just the usual four-point closed string
Veneziano amplitude which has to be integrated over internal momenta
of the $V_5$-vertices, i.e. over two out of three independent momenta.
The straightforward computation of the four-point amplitude and 
the integration over the $V_5$ momenta has been performed in ~\refs{\myself}
and the answer is given by:
\eqn\grav{\eqalign{A_{\lambda_0^2}\sim
\lambda_0^2{(p_1^||)^4}{log}(p_1^{||})^2\int
{d^2}w{{{log(|log|w||)}+log(|log|1-w||)}\over{|1-w|^4}}}}
where $p_1^{||}$ is the longitudinal projection of the dilaton
momentum to four longitudinal directions;
$(p^||)^2=p_\alpha{p}^\alpha$. It is remarkable that the amplitude
(8) depends exclusively on four-dimensional longitudinal projection
of the dilaton momentum; up to normalization it  has the same form
as the two-point correlator $<Tr{F^2}(p_1^{||})Tr{F^2}(-p_1^{||})>$
in the $N=4$ super Yang-Mills theory in $D=4$, computed
in the  approximation of dilaton s-wave ~\refs{\ampf}.
 Fourier transforming  the amplitude (8), one recovers
the well-known expression for the two-point
amplitude in the $N=4$ $D=4$  SYM theory in the four-dimensional
coordinate space:
$Tr{F^2}(x)Tr{F^2}(y)\sim{1\over{|x-y|^8}}$.
Furthermore, the momentum structure of  amplitudes with
more $V_5$ insertions agrees with the form of the
$<Tr{F^2}Tr{F^2}>$ correlators computed at higher values
of the dilaton angular momentum; in other words,
expansion in the $\lambda_0$ parameter in the
brane-like  sigma-model (2)  accounts for
higher partial waves of the dilaton field in the 
$AdS_5\times{S^5}$ supergravity.
Proceeding similarly, one can in principle  compute higher point
correlation functions  from the generating functional
(5) to show their agreement with the known expressions
for 3 and 4-point correlators in the $N=4,D=4$ SYM theory.
Regarding the four-point functions, one can recover the 
logarithmic singularities in the position space, observed previously
in the $AdS_5$  supergravity computations 
~\refs{\dhoker, \myselfin}.

In this paper we will explore the mechanism of
the dynamical compactication of flat  ten-dimensional
space-time on $AdS_5\times{S^5}$  caused by the presence
of potential term with the $V_5$ vertex in the brane-like sigma-model
(9). Our explanation of this mechanism is that the compactification 
on $AdS_5\times{S^5}$ occurs
as a result of  certain very special non-Markovian stochastic process.
 Namely, we will show that
 the $V_5$ insertion in the sigma-model
(5) has a meaning of a ``random force'' term with the $V_5$-operator
playing the role of a non-Markovian stochastic noise, 
which correlations are determined
by the worldsheet  beta-function associated with the $V_5$ vertex.
We will see that  the presence of the $V_5$-term in the
sigma-model (2) leads to Langevin-type
equation in the low-energy effective theory.
The role of the stochastic time variable in this Langevin equation
will be played by the worldsheet cutoff.
This means that the $V_5$-term effectively describes the interaction of
critical NSR string theory  with some infinite thermal reservoir,
breaking thermal equilibrium and causing the theory to drift 
stochastically, behaving like a brownian-type object.
Moreover, the reservoir itself is not in the equilibrium, which is
reflected in the non-Markovian properties of the $V_5$-noise.
Therefore the $V_5$-noise is not the usual white noise 
but it propagates in the stochastic time. Such a
 propagation is defined by the worldsheet correlators of the
$V_5$-vertices. As a result of the
 reservoir evolution,  the state of thermal equilibrium, reached in the limit
of infinite stochastic time in the Langevin equations, is $different$
from the initial  equilibrium state.
In other words, the overall scenario is the following:
in the beginning we have a field theory in $flat$ space time,
corresponding to the low-energy limit of usual NSR string theory.
This corresponds to the situation when we are  very far from the
$D3$-brane, corresponding to the $V_5$-insertion.
Next, the $D3$-brane, or
the non-Markovian $V_5$-noise appears in the picture.
This leads to the appropriate Langevin-type equations
for the dilaton, associated
with corresponding
modified Fokker-Planck equation ( this FP equation should be
modified due  to non-trivial correlations of the noise).
Next, we shall find that the distribution solving this modified
FP equation is given by the exponent of the $AdS_5$ supergravity
Hamiltonian, living on a constant time slice.
Evolution in the stochastic time corresponds to the flow in the direction
transverse to the $D3$-brane worldvolume.
Finally, in the limit of the infinite stochastic time we reach
$another$  equilibrium limit, corresponding to the theory living
in the $D3$-brane worldvolume, i.e. the four-dimensional
gauge theory.
Of course, such a scenario also implies
the correspondence between radial AdS coordinate and the stochastic time
which in turn corresponds to the worldsheet cutoff in the model (2).
This interpretation
 also implies that the
expansion in the $\lambda_0$ parameter in the brane-like sigma-model (9)
 corresponds to the the ``stochastic
diagrams'' which have to be averaged over the white
 noise distribution. It is possible that the consequence
of such an averaging is reflected in the numerical factor
given by  the integral over the worldsheet coordinate $w$ appearing 
in the amplitude (8), the only worldsheet integration left after
fixing Koba-Nielsen's measure in the four-point correlator.
  Though the last suggestion seems rather heuristic
and we will not provide any evidence for it in this paper,
 it nevertheless  seems very tempting to ask
whether the double logarithm appearing in the integral over $w$ in 
(8) has anything to do with the one appearing in the well-known
law of the iterated logarithm which limits the maximum distance 
covered by a Brownian particle for a time interval t:
$|x(t)-x(0)|<const\times{\sqrt{t}}log|log|t||$.
Indeed, in this paper we will point out connections
between the radial worldsheet coordinate (more precisely, the 
worldsheet cutoff), the stochastic time parameter and the radial
$AdS_5$ coordinate.
With all these intuitive ideas in mind, 
in the following section we will proceed to
deriving the Langevin equation from the brane-like sigma-model (2)
and the corresponding Fokker-Planck equation.

\vfill\eject

\centerline{\bf Langevin and Fokker-Planck equations from the
brane-like sigma-model}

Consider expansion of the generating functional (5) in
$\varphi(X)$ and $\lambda_0$.
We get:
\eqn\grav{\eqalign{Z(\lambda_0,\varphi)\sim
\int{D}\lbrack{X}\rbrack{D}\lbrack\psi\rbrack{D}\lbrack{ghosts}\rbrack
e^{-S_0^{NSR}}{\lbrace}1+\int{d^{10}p}\varphi(p)
\int{d^2z}V_\varphi(p,z,{\bar{z}})\cr+{1\over2}\int{d^{10}p_1}
\int{d^{10}p_2}\varphi(p_1)\varphi(p_2)\int{d^2}{z_1}
\int{d^2}{z_2}V_\varphi(p_1,z_1,{\bar{z_1}})V_\varphi(p_2,z_2,{\bar{z_2}})
\cr+\lambda_0\int{{d^4{k^{||}}}\over{{k^{||}}^4}}\int{d^2w}
V_5(k^{||},w,{\bar{w}})\cr+{1\over2}\lambda_0^2
\int{{d^4{k_1^{||}}}\over{{k_1^{||}}^4}}
\int{{d^4{k_2^{||}}}\over{{k_2^{||}}^4}}
\int{d^2w_1}\int{d^2w_2}V_5(k_1^{||},w_1,{\bar{w_1}})
V_5({k_2^{||}},w_2,{\bar{w_2}})\cr+{{\lambda_0^3}\over6}
\int{{d^4{k_1^{||}}}\over{{k_1^{||}}^4}}
\int{{d^4{k_2^{||}}}\over{{k_2^{||}}^4}}
\int{{d^4{k_3^{||}}}\over{{k_3^{||}}^4}}
\int{d^2w_1}\int{d^2w_2}\int{d^2w_3}
V_5(k_1^{||},w_1,{\bar{w_1}})\cr\times{V_5}(k_2^{||},w_2,{\bar{w_2}})
V_5(k_3^{||},w_3,{\bar{w_3}})+...\rbrace{f(\Gamma,N)}}}
where $S_0^{NSR}$ is the free part of NSR 
superstring action in flat space-time and 
we only retained the terms relevant to our discussion.
Consider first the term proportional to the square of the dilaton
in the expansion (9). The O.P.E. between two dilaton vertex operators
is given by
\eqn\grav{\eqalign{L_\varphi=\int{d^{10}p_1}\int{d^{10}p_2}
\int{d^2z_1}\int{d^2z_2}
V_\varphi(z_1,\bar{z_1},p_1)
V_\varphi(z_2,{\bar{z_2}},p_2)\sim\cr\sim\int{d^{10}p_1}\int{d^{10}p_2}
\int{d^2z_1}\int{d^2z_2}\lbrack
{{C_\varphi(p_1,p_2)}\over{|z_1-z_2|^2}}V_\varphi
({1\over2}(z_1+z_2),{1\over2}({\bar{z_1}}+
{\bar{z_2}}),p_1+p_2)+...\rbrack\cr
C_\varphi(p_1,p_2)\sim{(p_1-p_2)^2}}}
Here $C_\varphi(p_1,p_2)$ is the structure constant
and we have dropped all the terms not contributing to
the dilaton's beta-function.
The worldsheet integrals in (9), (10) are divergent as
the logarithmic
divergence originates from the region $|z_1-z_2|\sim\Lambda$
where $\Lambda$ is the worldsheet cutoff.
The relevant cutoff dependent part of the O.P.E. in (10) is then
given by
\eqn\grav{\eqalign{L_\varphi\sim{log{\Lambda}}\int{d^{10}p_1}
\int{d^{10}p_2}\int{d^2z}
C_\varphi{\varphi(p_1)\varphi(p_2)}{V_\varphi}(z,{\bar{z}},p_1+p_2)\cr
\cr=log\Lambda\int{d^{10}}p\int{d^2z}V_\varphi(p,z,{\bar{z}})\int{d^{10}q}
C_\varphi(q)\varphi({{p+q}\over2})\varphi({{p-q}\over2})}}
where we have  changed the variables:
\eqn\grav{\eqalign{z=z_1+z_2\cr
w=z_1-z_2\cr
p=p_1+p_2\cr
q=p_1-p_2}}

Combining this with the linear dilaton term in (9), it is easy to see that
in the absence of the $V_5$ term the  renormalization
of the space-time dilaton field necessary to remove the 
logarithmic divergence in the partition function (9) would be given by
\eqn\lowen{\varphi(p)\rightarrow{\varphi(p)
-log\Lambda\int{d^{10}q}C_\varphi(q)\varphi({{p+q}\over2})
\varphi({{p-q}\over2})}}
provided that one does not turn on the graviton excitation.
The condition for the worldsheet conformal invariance
leads then to low energy effective equations of motion for the  dilaton
field in flat space-time:
\eqn\lowen{\beta_\varphi\equiv{{d\varphi}\over{d(log{\Lambda})}}
=\int{d^{10}q}C_\varphi(q){\varphi({{p+q}\over2})\varphi({{p-q}\over2})}=0}
The presence of the $V_5$ vertex, however, modifies the dilaton
beta-function. The relevant part of the operator algebra is given by:
\eqn\lowen{V_5(w_1,\bar{w_1},k_1^{||})V_5(w_2,\bar{w_2},k_2^{||})
\sim{{C_{5\varphi}(k_1^{||},k_2^{||})}\over{|w_1-w_2|^2}}
V_\varphi({1\over2}(w_1+w_2),{1\over2}({\bar{w_1}}+{\bar{w_2}});
k_1^{||}+k_2^{||})}
where $C_{5\varphi}\sim(k_1^{||}+k_2^{||})^4$  is again the appropriate
structure constant; it is obtained easily by  computing the three-point
correlator of two $V_5$ states with the dilaton on the sphere.
It is remarkable that this structure constant is proportional
to the fourth power in the momentum and has no contributions quadratic
in $p$, unlike the case of the graviton.
Proceeding precisely as above and substituting
the structure constant $C_{5\varphi}$ we find that
the  logarithmically divergent contributions relevant to the dilaton's beta 
function are given by: 
\eqn\grav{\eqalign{
Z(\lambda_0,\varphi)\sim
\int{D}\lbrack{X}\rbrack{D}\lbrack\psi\rbrack{D}\lbrack{ghosts}\rbrack
e^{-S_0^{NSR}}{\lbrace}1+\int{d^{10}p}\int{d^2z}V_\varphi(z,{\bar{z}},p)
\lbrace\varphi(p)\cr+log\Lambda\lbrack
\int{d^{10}q}C_\varphi(q)\varphi({{p-q}\over2})\varphi({{p+q}\over2})
\cr+\lambda_0^2\int{d^{4}k_1^{||}}{{1}\over
{(p^{||}-k_1^{||})^4}}
(1+\lambda_0\int{{d^4k_2^{||}}\over{k_2^{||}}}\int_\Lambda{d^2w}
V_5(w,\bar{w},k_2^{||}))\rbrack\rbrace\rbrace}}
where the integral over $w$ is taken outside the two-dimensional
cutoff region.
To analyze this cutoff dependence, we write
$w=re^{i\alpha}$ and
\eqn\grav{\eqalign{\int_\Lambda{d^2w}V_5(w,\bar{w},k^{||})=
\int_0^{2\pi}{d\alpha}\int_\Lambda^{\infty}{dr}rV_5(r,\alpha,k^{||})=
\int_0^{2\pi}d\alpha\int_0^\infty{dr}rV_5(r+\Lambda,\alpha,k^{||})}}
The integral over $k_1^{||}$in (16) is 
taken over the on-shell surface;
it is computed by introducing the exponential regulator:
$\int{{d^4{k^{||}}}\over{{k^{||}}^4}}\rightarrow
\int{{d^4{k^{||}}e^{i\alpha{k}}}\over{{k^{||}}^4}}$
and then taking the limit $\alpha\rightarrow{0}$
The value of this integral is equal to
$1+log{\alpha^2}$ which gives just 1 after the regularization.
Therefore the dilaton's beta-function equation following from the partition
function expansion is
given by
\eqn\lowen{
{{d\varphi(p)}\over{d(log\Lambda)}}=
-\int{d^{10}q}C_\varphi(q)\varphi({{p-q}\over2})\varphi({{p+q}\over2})
+\eta_{5}(p^{||},\Lambda)}
where
\eqn\lowen{\eta_{5}(p^{||},\Lambda)\equiv
-\lambda_0^2
(1+\lambda_0\int{d^4k_2^{||}}
\int_0^{2\pi}d\alpha\int_0^\infty{dr}rV_5(r+\Lambda,\alpha,k^{||}))}
The 
dilaton beta-function equation (18) has the form of the Langevin equation
(1), with the role of the stochastic noise term being played by 
the truncated worldsheet integral of the $V_5$-vertex (19).
The logarithm of the
worldsheet  cutoff parameter plays the role of the stochastic
time in the  Langevin equation (18).
The noise (19) is non-Markovian and it is generated by
 the $V_5$ operator (3), as was already noted above.
Prior to computing  the correlators
of the $V_5$-noise (19), let us derive the 
Fokker-Planck equation associated with the Langevin equation (18).
It should be stressed that
in our derivation we will have to take into account that the noise is 
non-Markovian, i.e. its correlators depend non-trivially on 
the stochastic time.
Let us assume the two-point noise correlation function is given by
\eqn\lowen{<\eta_{5}(p_1^{||},\tau_1)\eta_{5}(p_2^{||},\tau_2)>
{\equiv}G(p_1^{||},p_2^{||},\tau_1,\tau_2)}
where $\tau\sim{log\Lambda}$ and
the noise Green's function $G(p_1^{||},p_2^{||},\tau_1,\tau_2)$
is  determined by properly regularized worldsheet correlators
of the $V_5$ vertices in (19). The precise form of this Green's
function will be computed below.
Let us furthermore  assume that the Green's function (19) corresponds
to some local effective action $I(\eta_{V_5})$ which determines
the propagation
of the noise in  stochastic time:
\eqn\lowen{I(\eta)\sim\int{d^4p{\int}d\tau(\eta_{5}{K}\eta_5+V(\eta_5))}}
where $K$ is some differential operator and $V$ is the potential.
To derive the Fokker-Planck equation, associated with the non-Markovian
Langevin equation (18), consider an arbitrary $\tau$-dependent
functional of the dilaton field $f(\varphi(p,\tau),\tau)$
Next, consider the average of the time derivative of this functional,
taking the Langevin equation (18) into account:
\eqn\grav{\eqalign{<{{df(\varphi,\tau)}\over{d\tau}}>=
\int{d^4p}<{{\delta{f}(\varphi,\tau)}\over{\delta{\varphi(p,\tau)}}}
{{d\varphi}\over{d\tau}}>\cr=
\int{d^4p}\lbrace
-<{{\delta{f}(\varphi,\tau)}\over{\delta{\varphi(p,\tau)}}}
{{\delta{S_E}}\over{\delta\varphi}}>
+<{{\delta{f}(\varphi,\tau)}\over{\delta{\varphi(p,\tau)}}}\eta_5
(-p,\tau)>\rbrace}}
where $S_E$ is the dilaton low energy effective action in
flat space-time (i.e. in the absence of the $V_5$-term and the graviton).
Integrating by parts, we can write
the left-hand side of (22) as
\eqn\lowen{<{{d{f}(\varphi,\tau)}\over{d{\tau}}}>
=\int{D\varphi}P_{FP}(\varphi,\tau){{d{f}(\varphi,\tau)}
\over{d{\tau}}}=-\int{D\varphi}{{dP_{FP}(\varphi,\tau)}\over{d\tau}}
f(\varphi,\tau)}
where $P_{FP}(\varphi,\tau)$ is five-dimensional
Fokker-Planck distribution associated with the Langevin
equation (18). Next,
the first correlator  on the right hand side of (22),
involving $S_E$ may be written as
\eqn\grav{\eqalign{<{{\delta{f}(\varphi,\tau)}\over{\delta{\varphi(p,\tau)}}}
{{\delta{S_E}}\over{\delta\varphi}}>=
\int{d^4p}\int{D\varphi}P_{FP}(\varphi,\tau)
<{{\delta{f}(\varphi,\tau)}\over{\delta{\varphi(p,\tau)}}}
{{\delta{S_E}}\over{\delta\varphi}}>\cr=
-\int{D\varphi}f(\varphi,\tau)\int{d^4p}\int{d^4q}
{{\delta}\over{\delta\varphi
(p,\tau)}}\lbrack{{\delta{S_E}}\over{\delta\varphi(q)}}P_{FP}(\varphi,\tau)
\rbrack}}
given that partial integration was again  performed.
Finally, consider the second term in the right hand side
of (22).
We have:
\eqn\grav{\eqalign{\int{d^4}p
<{{\delta{f}(\varphi,\tau)}\over{\delta{\varphi(p,\tau)}}}\eta_5(-p,\tau)>=
\int{D\eta_5}\int{d^4p}e^{-I(\eta_5)}
{{\delta{f}(\varphi,\tau)}\over{\delta{\varphi(p,\tau)}}}\eta_5(-p,\tau)\cr
=lim_{j\rightarrow{0}}\int{d^4p}
{\delta\over{\delta{j(-p,\tau)}}}\int{D\eta_5}
e^{-I(\eta_5)+\int{d^4q}{\int}{dt}j(q,t)\eta_5(q,t)}
{{\delta{f}}\over{\delta\varphi}}
(p,\tau)}}
where we introduced infinitezimal source term in the last expression.
Let us perform the transformation:
\eqn\lowen{\eta_5\rightarrow\eta_0+{\tilde\eta}}
where $\eta_0$ is the solution of equation ${{\delta{I}}\over{\delta\eta}}
=-j$
The functional integral (25) becomes
\eqn\grav{\eqalign
{lim_{j\rightarrow{0}}{\delta\over{\delta{j}}}
\int{d^4}p\int{D\tilde{\eta}}
e^{-{\lbrack}I({\tilde\eta})+\int{d^4k{\int}d\tau}
{{\delta{I({\tilde\eta})}\over{\delta{\tilde\eta}}}}\eta_0(k,\tau)+...\rbrack}
\times\lbrace{{\delta{f}}\over{\delta\varphi}}(p,\tau)\cr+\int{d^4q\int{dt}}
\eta_0(q,t){\delta\over{\delta{\tilde\eta}(-q,t)}}
({{\delta{f}}\over{\delta\varphi}}(p,\tau))+O(\eta_0^2)\rbrace}}
In case when $I(\eta_5)=\int{\eta_5}{K}{\eta_5}$,
 where K is some self-adjoint operator,
we have $\eta_0(q,\tau)=-\int{d^4kdt}G(q,\tau,k,t)j(-k,t)$
where $G$ is the noise two-point function.
Substituting into (27) we get:
\eqn\grav{\eqalign{{lim_{j\rightarrow{0}}}{\delta\over{\delta{j}}}\int{d^4}p
\int{D{\tilde{\eta}}}
e^{-{\lbrack}I(\tilde{\eta})+\int{d^4}q{\int}d\tau\int{d^4}k{\int}dt
j(q,\tau)G(-q,-k,\tau,t)j(k,t)\rbrack}
\lbrace{{\delta{f}}\over{\delta\varphi}}(p,\tau)\cr+
\int{d^4q}\int{dt}
{\delta\over{\delta{\tilde\eta}(q,t)}}
({{\delta{f}}\over{\delta\varphi}}(p,\tau))j(q,t)G(-p,\tau,-q,t)
+O(j^2)\rbrace\cr=\int{D{\tilde{\eta}}}e^{-I({\tilde{\eta}})}
\int{d^4q}{\int}dt\int{d^4p}G(-p,\tau,-q,t)
{{\delta\over{\delta{\tilde{\eta}}(q,t)}}}
{{\delta{f}}\over{\delta\varphi}}(p,\tau)\cr=
\int{d^4q}{\int}dt\int{d^4p}
G(-p,\tau,-q,t)<{\delta\over{\delta\eta(q,t)}}
{{{\delta{f}}\over{\delta\varphi(p,\tau)}}}>\cr=
\int{d^4q}dt\int{d^4p}
G(-p,\tau,-q,t)<{\delta\over{\delta\varphi(q,t)}}
{{{\delta{f}}\over{\delta\varphi(p,\tau)}}}>}}
where we used 
the identity
\eqn\lowen{{\delta\over{\delta\eta(q,t)}}={\delta\over{\delta\varphi
(q,t)}}}
To derive it, note that 
 the general solution to the Langevin
equation (18) is given by:
\eqn\lowen{\varphi(k,\tau)=\int{dt}e^{-k^2(t-\tau)}
\theta(t-\tau)\eta(k,t)}
for which 
${{\delta\varphi(k,\tau)}\over{\delta\eta(k,\tau)}}=\theta(0)=1$.
~\refs{\huffel}
It is easy to see that in cases when $I(\eta)$ is not quadratic,
the relation (28) is still true in the WKB limit.
Next, writing the average (28) in terms of the functional
integral in $\varphi$ and integrating by parts as in (23) and
(24) we obtain
\eqn\grav{\eqalign{
\int{d^4p}<{{\delta{f}(\varphi,\tau)}\over{\delta\varphi(p,\tau)}}>\cr=
\int{D\varphi}f(\varphi,\tau)
\int{d^4p}\int{d^4q}{\int}dt{\delta\over{\delta\varphi(q,t)}}
G(-p,\tau,-q,t){\delta\over{\delta\varphi(p,\tau)}}P_{FP}(\varphi,\tau)}}
Finally, putting together (23), (24), (31) and using the Langevin
equation (18) we obtain the  following Fokker-Planck equation
for the non-Markovian stochastic process (18), (19):

\eqn\grav{\eqalign{{{d{P_{FP}}}\over{d\tau}}=
\int{d^4p}\int{d^4q}{{\delta}\over{\delta\varphi
(q,\tau)}}\lbrack{{\delta{S_E}}\over{\delta\varphi(p)}}P_{FP}(\varphi,\tau)
\rbrack\cr-
\int{d^4p}\int{d^4q}{\int}dt{\delta\over{\delta\varphi(q,t)}}
G(-p,\tau,-q,t){\delta\over{\delta\varphi(p,\tau)}}P_{FP}(\varphi,\tau)}}
To complete our derivation we need to evaluate the two-point noise
correlator $G(p,q,\tau,t)$ introduced in (20), 
which is determined by the worldsheet correlations
of the $V_5$ vertex in (19). Firstly, let us compute
the correlator $<\int_{\Lambda_1}{d^2z}V_5(z,\bar{z})\int_{\Lambda_2}
{d^2w}V_5(w,{\bar{w}})>$ of the $V_5$-operators entering the
definition (19) of the noise.
We have:
\eqn\grav{\eqalign{<{\Gamma^4}
\int_{\Lambda_1}{d^2z}V_5(z,\bar{z})\int_{\Lambda_2}
{d^2w}V_5(w,{\bar{w}})>=6\int_{\Lambda_1}{d^2z}\int_{\Lambda_2}
{d^2w}{{1\over{{|z-w|^4}}}}}}
where the fourth power of picture-changing operator $\Gamma^4$
has been introduced in the correlator to insure the cancellation
of ghost number anomaly on a sphere.
The integrals are divergent and their regularized dependence
on $\Lambda_1$, $\Lambda_2$ determines the form of noise
correlations in stochastic time.                            
Writing $z=r_1e^{i\alpha_1},w=r_2e^{i\alpha_2}$
we have:
\eqn\grav{\eqalign{
<\int_{\Lambda_1}{d^2z}V_5(z,\bar{z})\int_{\Lambda_2}
{d^2w}V_5(w,{\bar{w}})>\cr=
6\int_{\Lambda_1}^{\infty}dr_1{r_1}{\int_0^{2\pi}}d\alpha_1
\int_{\Lambda_2}^{\infty}dr_2{r_2}{\int_0^{2\pi}}d\alpha_2
{1\over{(r_1^2+r_2^2-2r_1r_2cos(\alpha_1-\alpha_2))^2}}\cr=
3\pi^2{\int_{\Lambda_1}^\infty}d{r_1^2}{\int_{\Lambda_2}^{\infty}}d{r_2^2}
\lbrace{{1\over{(r_1^2-r_2^2)^2}}}+{{2r_2^2}\over{(r_1^2-r_2^2)^3}}\rbrace}}
The regularized value of the of the first integral gives:
\eqn\lowen{{\int_{\Lambda_1}^\infty}d{r_1^2}
{\int_{\Lambda_2}^{\infty}}d{r_2^2}
{1\over{(r_1^2-r_2^2)^2}}=log(\Lambda_1^2-\Lambda_2^2)}
while the second one is given by
\eqn\lowen{
{\int_{\Lambda_1}^\infty}d{r_1^2}{\int_{\Lambda_2}^{\infty}}d{r_2^2}
{{2r_2^2}\over{(r_1^2-r_2^2)^3}}={{\Lambda_2^2}\over{\Lambda_1^2-\Lambda_2^2}}
-log(\Lambda_1^2-\Lambda_2^2)}
Adding these contributions together we see that the logarithmic terms
cancel. Recalling that $log\Lambda=\tau$ where $\tau$ is stochastic time
we get the following expression for the non-Markovian noise correlator:
\eqn\grav{\eqalign{<\eta_5(p^{||},\tau_1)\eta_5(q^{||},\tau_2)>
\sim
{\lambda_0^4}(1+
{{\lambda_0^2}\over{e^{2(\tau_1-\tau_2)}-1}})}}
Substituting this expression into the Green's function in (32)
concludes our derivation of the Fokker-Planck equation associated with
the non-Markovian stochastic process (18) in the brane-like sigma-model.
In the following section we will study the ansatz solutions to this
equation showing their relevance to the $AdS_5$ supergravity
Hamiltonian.

\vfill\eject

\centerline{\bf $AdS_5$ Supergravity as a solution of the 
Fokker-Planck Equation}
Consider the bosonic part of the action for the $AdS_5$ supergravity:
\eqn\lowen{S_{AdS_5}\sim{N^2}\int{d\lambda}{d^4x}{1\over{\lambda^3}}
(\partial_{\lambda}\varphi\partial_{\lambda}\varphi+
\partial_\mu\varphi\partial^\mu\varphi)}
where $\varphi(\lambda,x)$ is the dilaton, $(\lambda, x^\mu)$
are the transverse  and longitudinal $AdS_5$ coordinates ( $\mu=0,...3$).
The gauge-theoretic $N$ parameter entering (38) is related to
the $AdS_5$ radius $R$ and ten-dimensional gravitational constant 
$\kappa$ through ${{R^8}\over{\kappa^2}}\sim{N^2}$.
As is easy to check, the action (38) corresponds to
the $AdS_5$ metric with normalization
\eqn\lowen{ds^2={{N^{4\over3}}\over{\lambda^2}}
(d\lambda{d\lambda}+dx_\mu{dx^\mu})}
Substituting $t={N^{2\over3}}log\lambda$ we can write the metric as
\eqn\lowen{ds^2=dtdt+{N^{4\over3}}e^{-2t{N^{-{2\over3}}}}dx_\mu{dx^\mu}}
In the new coordinates corresponding  to the temporal gauge,
 the  gravity action becomes:
\eqn\lowen{S_{AdS_5}={N^{8\over3}}\int{dt}{d^4x}e^{-4t{{N^{-2\over3}}}}
(\partial_t\varphi
\partial_t\varphi+{{N^{-{4\over3}}}}
e^{2t{N^{2\over3}}}\partial_\mu\varphi\partial_\mu\varphi)}
The conjugate mimentum $\pi$, computed at a constant ``time''  t
is given by
\eqn\lowen{\pi={1\over{\sqrt{g}}}{{\delta{S_{AdS_5}}}\over{\delta\varphi}}=
{\partial_t}\varphi}
The corresponding ADM-type Hamiltonian is given by ~\refs{\verlinde, \verlind}
\eqn\lowen{H_{AdS_5}(t)=\int{d^4x}{\sqrt{g}}({1\over2}\pi^2+L)=
{N^{8\over3}}\int{d^4x}e^{-4t{N^{-{2\over3}}}}
({1\over2}\partial_t\varphi\partial_t\varphi
+{N^{-{4\over3}}}e^{2tN^{-{2\over3}}}\partial_\mu\varphi\partial^\mu\varphi)}
where $L=\partial_\mu\phi\partial^\mu\phi$ is the local Lagrangian density.
In the $(\lambda, x)$ coordinates this ``equal time slice''
Hamiltonian is given by:
\eqn\lowen{H_{AdS_5}(\lambda)={N^{4\over3}}\int{d^4x}{1\over{\lambda^2}}
({1\over2}\partial_{\lambda}\varphi\partial_{\lambda}
\varphi+\partial_\mu\varphi
\partial^\mu\varphi)}
We shall also need the expression for this Hamiltonian in the $({\rho},x)$
coordinates with ${\rho}=\lambda^2$:
\eqn\lowen{H_{AdS_5}(\rho)={N^{4\over3}}
\int{d^4x}(2\partial_{\rho}\varphi\partial_{\rho}
\varphi
+{1\over{\rho}}\partial_\mu\varphi\partial^\mu\varphi)}
Our strategy now is to  relate the radial $AdS_5$
coordinate to the stochastic time parameter.
This will allow us to interpret  the Hamiltonian
$H_{AdS_5}$ as a logarithm of the Fokker-Planck distribution
$P_{FP}\sim{e^{-H_{FP}}}$ solving the Fokker-Planck equation (32).
However, an important remark should be made here.
We know  from the holography principle  that the radial 
${{\rho}}$ coordinate
 of $AdS_5$ in (45) corresponds to the gauge theory cutoff;
therefore if one tries to interpret the AdS/CFT correspondence
in terms of the stochastic quantization and the radial $AdS$ coordinate
as a stochastic time, it is the gauge theoretic cutoff that should
enter the Langevin and Fokker-Planck equations solved by the
$AdS_5$ supergravity Hamiltonian (43), (44). 
At the same time, however, the stochastic variable in the Langevin
equation (18) corresponds to the worldsheet cutoff in string theory.
It is clear that the cutoffs in gauge and string theories are not the same;
 they are related to each other in some non-trivial way.
In fact, the knowledge of relation between
the string theoretic cutoff to the gauge-theoretic one
is necessary to make sense of any possible connections
between strings and Yang-Mills theories.
In other words, the $AdS_5$
gravity Hamiltonian (43), (44) must be a solution
to the the Fokker-Planck equation (32) expressed in terms
of the gauge theory cutoff (stochastic time) rather than
the string-theoretic one.
The crucial question is what precisely is the relation between
$\rho$ and $\Lambda_{string}\equiv{\Lambda}$.
Below we will show that choosing
\eqn\lowen{\rho=log{\Lambda}}
and expressing the Langevin equation (18), (19) in terms
of $\rho$ (as well as the associate Fokker-Planck equation)
one can see that $P_{FP}=e^{-H_{AdS_5}}$ is the solution.
Consider the Langevin equation (18). Substituting (46) for $\Lambda$
we have:
\eqn\lowen{{}{{d\varphi(p)}\over{d\rho}}=
-{{\delta{S_E}}\over{\delta\varphi}}+\eta_5(p,e^{\rho})}
where 
\eqn\lowen{\eta_5(p,e^{2\rho})\sim
{{\lambda_0^2}}(1+\lambda_0\int_{e^{\rho}}d^2z
V_5(z,{\bar{z}}))}
and the worldsheet integral is cut off at the scale $e^{2\rho}$.
To cut the theory at the size of $\rho$ corresponding to the
the gauge theory cutoff  we have to  conformally transform
$u=log{z}$. Using the conformal covariance
of the $V_5$ vertex operator we obtain
\eqn\lowen{\eta_5(p,\rho)\sim
{{\lambda_0^2}}(1+\lambda_0\int_{\rho}{d^2u}
V_5(u,{\bar{u}}))}
The  Fokker-Planck equation associated with
the stochastic process (47) is then given by:
\eqn\grav{\eqalign{{{d{P_{FP}}}\over{d\rho}}=
\int{d^4p}\int{d^4q}{{\delta}\over{\delta\varphi
(p,\rho)}}
\lbrack{{\delta{S_E}}\over{\delta\varphi(q,\rho)}}P_{FP}(\varphi,\rho)
\rbrack\cr
-\int{d^4p}\int{d^4q}\int{d\omega}{\delta\over{\delta\varphi(q,\omega)}}
{G(p,\rho,q,\omega)}
{\delta\over{\delta\varphi(p,\rho)}}P_{FP}(\varphi,\rho)}}
where 
\eqn\lowen{G(p,\rho,q,\omega)=G(-p,\rho,-q,\omega)={{\lambda_0^4}}
(1+\lambda_0^2{{\omega^2}\over{\omega^2-\rho^2}})},
i.e. it has the same form as
the correlator (37) expressed in terms of $\Lambda_1$,
$\Lambda_2$. 
The integral over $\omega$ is taken from minus infinity to infinity.
We are now ready to check that the Fokker-Planck 
equation (50) with  noise propagator (51) is satisfied by the
distribution with Hamiltonian (45).
It is convenient to write the Hamiltonian (45) as
\eqn\lowen{H_{AdS_5}(\rho)=H_1(\rho)+H_2(\rho)}
where
\eqn\grav{\eqalign{
H_1(\rho)=2{N^{4\over3}}\int{d^4x}\partial_\rho\varphi\partial_\rho\varphi
=2{N^{4\over3}}
\int{d^4q}\partial_\rho\varphi(q,\rho)\partial_\rho\varphi(-q,\rho)\cr
H_2(\rho)={{N^{4\over3}}\over{\rho}}\int{d^4x}
\partial_\mu\varphi\partial^\mu
\varphi={{N^{4\over3}}\over{\rho}}\int{d^4q}
\varphi(q,\rho){q^2}\varphi(-q,\rho)}}
and the Fourier transform to the momentum space has been performed.
We shall use the ``constant time slice'' variational calculus
implying ${{\delta\varphi(q_1,\rho_1)}\over{\delta\varphi(q_2,\rho_2)}}
=\delta^{(4)}(q_1-q_2)\theta_{\rho_1-\rho_2}$
where $\theta_\rho$ is the function which is 1 at $\rho=0$ and
zero elsewhere. This function can be expressed in terms of 
the usual Heaviside step function as $\theta_\rho=\theta(\rho)\theta(-\rho)$
Note that the product of two Heaviside functions is a
well-defined object in the context of the Colombeau theory of multiplication
of distributions ~\refs{\colomb}
The useful identity for $\theta_\rho$ is
\eqn\lowen{\int_{-\infty}^{\infty}d\rho{{\theta_\rho}\over{\rho}}=1}
which can be easily checked using the integral representation
for the Heaviside function:
$\theta(\rho)=\int_{-\infty}^{\infty}dx{{e^{ix\rho}}\over{x}}$.
It is convenient to represent the r.h.s. of the Fokker-Planck equation
(50) (apart from the drift term)
 as ${\hat{I_1}+\hat{I_2}+...+\hat{I_6}}P_{FP}$ where
schematically
\eqn\grav{\eqalign{{\hat{I_1}}=\int{{\delta\over{\delta\varphi}}}G
{{\delta\over{\delta\varphi}}}H_1\cr
{\hat{I_2}}=\int{{\delta\over{\delta\varphi}}}G
{{\delta\over{\delta\varphi}}}H_2\cr
{\hat{I_3}}=\int{{\delta{H_1}}\over{\delta\varphi}}G
{{\delta{H_1}}\over{\delta\varphi}}\cr
{\hat{I_4}}=\int{{\delta{H_2}}\over{\delta\varphi}}G
{{\delta{H_2}}\over{\delta\varphi}}\cr
{\hat{I_5}}=\int{{\delta{H_1}}\over{\delta\varphi}}G
{{\delta{H_2}}\over{\delta\varphi}}\cr
{\hat{I_6}}=\int{{\delta{H_2}}\over{\delta\varphi}}G
{{\delta{H_1}}\over{\delta\varphi}}
}}
First, let us consider
\eqn\grav{\eqalign{{\hat{I_2}}P_{FP}(\rho)=\lbrace\int{d^4q}
\int{d^4p{d}\omega}{{\delta\over{\delta\varphi(q,\rho)}}}
G(p,\rho,q,\omega){{\delta\over{\delta\varphi(p,\omega)}}}H_2(\rho)\rbrace
P_{FP}(\rho)
\cr=\lbrace{\lambda_0^4}{N^{4\over3}}\int{d^4p}\int{d^4q}\delta^{(4)}(p-q)
\int{d\omega}{\theta_{\omega-\rho}}(1+
\lambda_0^2{{p^2}\over{\omega}}{{\omega^2}\over{\omega^2-\rho^2}})\rbrace
P_{FP}(\rho)}}
Using (54) we can evaluate the $\omega$ integral in (55)
and obtain:
\eqn\lowen{{\hat{I_1}}P_{FP}(\rho)=
{{{N^{4\over3}}\lambda_0^6}\over2}\int{d^4q}q^2{P_{FP}}(\rho)}
Next, consider the piece
\eqn\grav{\eqalign{{\hat{I_1}}P_{FP}(\rho)=\lbrace\int{d^4q}
\int{d^4p{d}\omega}{{\delta\over{\delta\varphi(q,\rho)}}}
G(p,\rho,q,\omega){{\delta\over{\delta\varphi(p,\omega)}}}H_1(\rho)\rbrace
P_{FP}(\rho)}}
To compute this piece, consider first of all the variation:
\eqn\grav{\eqalign{{{\delta{H_1}(\rho)}\over{\delta\varphi(q,\rho)}}
P_{FP}(\rho)
={N^{4\over3}}{\delta\over{\delta\varphi(q,\rho)}}\int{d^4p}\int{du}
\partial_u\varphi(p,u)\partial_u\varphi(-p,u)\delta(\rho-u)P_{FP}(u)\cr=
-{N^{4\over3}}{{\delta\over{\delta\varphi(-q,\rho)}}}
\int{d^4p}\int{du}\varphi(p,u)\partial_u\lbrack\delta(\rho-u)\partial_u\varphi
(-p,u)P_{FP}(u)\rbrack=\cr-{N^{4\over3}}
\int{du}\theta_{\rho-u}\partial_{u}\lbrack
\delta(\rho-u)\partial_u\varphi(-q,u)P_{FP}(u)\rbrack\cr=
-{N^{4\over3}}\int{du}\varphi(-q,u)\partial_u\lbrack\delta(\rho-u)\partial_u
\theta_{\rho-u}P_{FP}(u)\rbrack}}
Next, performing the second variation and integrating by parts we have:
\eqn\grav{\eqalign{{\hat{I_1}}P_{FP}(\rho)=
-{N^{4\over3}}\int{du}\int{d\omega}(\lambda_0^4+\lambda_0^6
{{\omega^2}\over{\omega^2-\rho^2}})\theta_{\omega-u}\partial_u
\lbrack\delta(\rho-u)\partial_u\theta_{\rho-u}P_{FP}(u)\rbrack\cr
={N^{4\over3}}\int{du}\int{d\omega}
(\lambda_0^4+{{\lambda_0^6\omega^2}\over{\omega^2-\rho^2}})
\partial_u\theta_{\omega-u}\delta(\rho-u)
\partial_u\theta_{\rho-u}P_{FP}(u)\cr=
-{N^{4\over3}}
\int{dw}(\lambda_0^4+{{\lambda_0^6\omega^2}\over{\omega^2-\rho^2}})
{\partial_\omega}J(\rho,\omega)}}
where we used 
$\partial_u\theta_{\omega-u}=-\partial_{\omega}\theta_{\omega-u}$
and denoted $J(\rho,\omega)=\int{du}\theta_{\omega-u}\delta(\rho-u)
\partial_{u}\theta_{\rho-u}P_{FP}(u)$.
Next, integrating by parts again and using the above
identity for the derivatives
of $\theta$ we obtain:
\eqn\grav{\eqalign{J(\rho,\omega)=
\int{du}
\partial_{u}\theta_{\rho-u}\theta_{\omega-u}\delta(\rho-u)P_{FP}(u)
=-\int{du}\partial_{\omega}\theta_{\rho-u}\theta_{\omega-u}\delta(\rho-u)
P_{FP}(u)\cr=
-\partial_\rho\int{du}\theta_{\rho-u}\theta_{\omega-u}\delta(\rho-u)
P_{FP}(u)+\int{du}\theta_{\rho-u}\theta_{\omega-u}(\delta(\rho-u))'
P_{FP}(u)\cr=
-\partial_\rho(\theta_{\omega-\rho}P_{FP}(\rho))-
\partial_\rho(\theta_{\omega-\rho}P_{FP}(\rho))
=-2\partial_\rho(\theta_{\omega-\rho}P_{FP}(\rho))}}
Substituting $J(\rho,\omega)$ into ${\hat{I_1}}P_{FP}(\rho)$
we obtain:
\eqn\grav{\eqalign{{\hat{I_1}}P_{FP}(\rho)=
-2{N^{4\over3}}\int{d\omega}
(\lambda_0^4+{{\lambda_0^6\omega^2}\over{\omega^2-\rho^2}})
\partial_{\omega}\partial_{\rho}(\theta_{\omega-\rho}P_{FP}(\rho))\cr=
-4{N^{4\over3}}
\lambda_0^6\rho^2\partial_{\rho}P_{FP}(\rho)\int{d\omega}
{{\omega\theta_{\omega-\rho}}\over{(\rho^2-\omega^2)^2}}+
4{N^{4\over3}}\lambda_0^6\rho^2{P_{FP}}(\rho)\int{d\omega}{{\omega
\partial_\omega{\theta_{\omega-\rho}}}\over{(\rho^2-\omega^2)^2}}\cr
=-4{N^{4\over3}}\lambda_0^6\rho^2\partial_\rho{P_{FP}}(\rho)
\partial_\rho({1\over{4\rho}})
-4{N^{4\over3}}
\lambda_0^6\rho^2{P_{FP}}(\rho)\int{d\omega}\theta_{\omega-\rho}
({1\over{(\rho^2-\omega^2)^2}}+{{4\omega^2}\over{(\rho^2-\omega^2)^3}})
\cr={N^{4\over3}}\lambda_0^6\partial_\rho{P_{FP}}(\rho)+{N^{4\over3}}
\lambda_0^6\rho^2(-{1\over{2\rho^3}}+{1\over{2\rho^3}})P_{FP}(\rho)
\lambda_0^6\partial_\rho{P_{FP}}(\rho)}}
To summarize,
\eqn\lowen{{\hat{I_1}}P_{FP}(\rho)={N^{4\over3}}
\lambda_0^6\partial{P_{FP}}(\rho)}
Next, consider
\eqn\grav{\eqalign{{\hat{I_3}}P_{FP}(\rho)=
\int{d^4p}\int{d^4q}\int{d\omega}
{{\delta{H_1}(\rho)}\over{\delta\varphi(\rho)}}
G(p,q,\rho,\omega){{\delta{H_1}(\rho)}\over{\delta\varphi(\omega)}}
P_{FP}(\rho)\cr=
\lambda_0^6{N^{8\over3}}
\int{d^4p}\int{d^4q}\int{d\omega}{{\omega^2}\over{\omega^2-\rho^2}}
\partial_\rho^2\varphi(p,\rho)\partial_\omega(\theta_{\omega-\rho}
\partial_\omega
\varphi(q,\rho))P_{FP}(\rho)\cr=\lambda_0^6{N^{8\over3}}
\int{d^4p}\int{d^4q}\lbrace\int{d\omega}{{\omega^2}\over{\omega^2-\rho^2}}
\theta_{\omega-\rho}\partial_\rho^2\varphi(p,\rho)
\partial_\omega^2\varphi(q,\omega)\cr-\int{d\omega}\theta_{\omega-\rho}
\partial_\omega({{\omega^2}\over{\omega^2-\rho^2}})\partial_\rho^2
\varphi(p,\rho)\partial_\omega\varphi(q,\omega)\rbrace{P_{FP}(\rho)}
\cr=
\lambda_0^6{N^{8\over3}}\int{d^4p}\int{d^4q}({
{\rho\over2}\partial_\rho^2\varphi(p,\rho)-{\rho\over2}
\partial_\rho^2\varphi(q,\rho)})\partial_\rho^2\varphi(p,\rho)P_{FP}(\rho)=0}}
The next piece is given by
\eqn\grav{\eqalign{{\hat{I_4}}P_{FP}(\rho)=
\int{d^4p}\int{d^4q}\int{d\omega}
(\lambda_0^4+\lambda_0^6{{\omega^2}\over{\omega^2-\rho^2}})
{{\delta{H_2(\rho)}}\over{\delta\varphi(p,\rho)}}
{{\delta{H_2(\rho)}}\over{\delta\varphi(q,\omega)}}P_{FP}(\rho)
\cr=\lambda_0^6{N^{8\over3}}
\int{d^4p}\int{d^4q}{{p^2q^2}\over{\rho}}\varphi(p,\rho)
\int{d\omega}{{\theta_{\omega-\rho}\omega\varphi(q,\omega)}
\over{(\omega-\rho)(\omega+\rho)}}P_{FP}(\rho)\cr=
{{\lambda_0^6{N^{8\over3}}}\over{2\rho}}\int{d^4p}\int{d^4q}p^2q^2
\varphi(p,\rho)\varphi(q,\rho){P_{FP}(\rho)}}}
Next,
\eqn\grav{\eqalign{{\hat{I_5}}P_{FP}(\rho)=
\int{d^4p}\int{d^4q}\int{d\omega}
(\lambda_0^4+\lambda_0^6{{\omega^2}\over{\omega^2-\rho^2}})
{{\delta{H_1}(\rho)}\over{\delta\varphi(\rho)}}
{{\delta{H_2}(\rho)\over{\delta\varphi(\omega)}}}P_{FP}(\rho)\cr=\lambda_0^6
{N^{8\over3}}
\int{d^4p}\int{d^4q}\int{d\omega}{{\omega^2}\over{\omega^2-\rho^2}}
\partial_\rho^2\varphi(p,\rho){{q^2}\over{\omega}}\varphi(q,\omega)
P_{FP}(\rho)\cr={{\lambda_0^6{N^{8\over3}}}\over2}
\int{d^4p}\int{d^4q}{q^2}\varphi(q,\rho)
{\partial_\rho^2}\varphi(p,\rho)P_{FP}(\rho)}}
At last,
\eqn\grav{\eqalign{{\hat{I_6}}P_{FP}(\rho)
=\int{d^4p}\int{d^4q}\int{d\omega}
(\lambda_0^4+\lambda_0^6{{\omega^2}\over{\omega^2-\rho^2}})
{{\delta{H_2(\rho)}}\over{\delta
\varphi(\rho)}}{{\delta{H_1(\rho)}}\over{\delta\varphi(\omega)}}P_{FP}(\rho)
\cr={{N^{8\over3}}\lambda_0^6}\int{d^4p}\int{d^4q}\int{d\omega}
{{p^2\varphi(p,\rho)}\over{\rho}}\partial_\omega
(\theta_{\omega-\rho}\partial_\omega\varphi(q,\omega))P_{FP}(\rho)=0}}
Finally, using (55)-(67) we find that substituting the $AdS_5$
supergravity Hamiltonian to
the ``kinetic'' part of the
r.h.s. of the Fokker-Planck equation (50) (i.e. the r.h.s. without
a drift term) is given by:
\eqn\grav{\eqalign{\int{d^4p}\int{d^4q}\int{d\omega}
{\delta\over{\delta\varphi(q,\omega)}}
G(p,\rho,q,\omega){\delta\over{\delta\varphi
(p,\rho)}}P_{FP}(\rho)={N^{4\over3}}\lambda_0^6
\lbrace{\partial\over{\partial\rho}}+
{1\over2}\int{d^4q}q^2\cr+{{N^{4\over3}}\over{2\rho}}\int{d^4p}\int{d^4q}
{p^2\varphi(p,\rho)q^2\varphi(q,\rho)}+
{{N^{4\over3}}\over2}\int{d^4p}\int{d^4q}{\partial_\rho^2\varphi(p,\rho)
q^2\varphi(q,\rho)}\rbrace{P_{FP}}(\rho)}}
The next step is to evaluate the drift term of the Fokker-Planck
equation (50). First of all, we have to point out the 
correct form and normaliaztion
 of the equilibrium action $S_E$. As the stochastic time
$\rho$ corresponds to the radial $AdS_5$ coordinate, orthogonal
to the worldvolume of the underlying D3-brane, the stochastic
process (18), (19) describes the evolution of an ``equal time slice''
gravity Hamiltonian as it drifts  from infinity towards the $D3$-brane
worldvolume. The state of thermal equilibrium, reached in the limit
of infinite  stochastic time, is described by the free field
scalar theory in four dimensions since in the region far from the D3-brane
the space-time is flat.
On the other hand, action in the drift term of the Langevin equation
(18) is the one of the scalar field theory in $ten$ dimensions,
i.e. the dilaton effective action (in the Einstein frame) in the
absence of the graviton mode and the $V_5$ term.
Note, however, that  truncation of the $\lambda_0$ expansion
in the brane-like sigma-model, performed in (9),
corresponds to the dilaton
s-wave approximation in the $AdS_5\times{S^5}$ picture, as
can be shown by considering sigma-model
 correlation functions and comparing them to those of the
$D=4$ super Yang-Mills theory ~refs{\myself}.
In the s-wave approximation the $AdS_5\times{S^5}$ supergravity is
reduced to the $AdS_5$ one with the action (38)  while the
ten-dimensional free field dilaton action entering the Langevin equation (18),
given by
\eqn\lowen{S_{10}\sim{1\over{\kappa^2}}\int{d^{10}x}\partial_\alpha\varphi
\partial^\alpha\varphi}
can be reduced to the five-dimensional one given by
\eqn\grav{\eqalign{S_5\sim{
{R^5}\over{\kappa^2}}\int{d^4x}\int{d\rho}(\partial_\mu\varphi\partial^\mu     
\varphi+\partial_\rho\varphi\partial_\rho\varphi)
\cr={{N^2}\over{R^3}}\int{d^4x}\int{d\rho}
(\partial_\mu\varphi\partial^\mu\varphi+\partial_\rho\varphi\partial^\rho
\varphi)}}
Of course, higher order $\lambda_0$ corrections 
(corresponding to dilaton partial waves with higher angular momenta)
modify the kinetic
part of the Fokker-Planck equation (50) and the full ten-dimensional
equilibrium action must then be used. 
 The ADM-type ``equal time slice'' four-dimensional Hamiltonian
may be constructed out of this action precisely as in the $AdS_5$ case;
in the limit $\rho\rightarrow{-}\infty$ it is given by
\eqn\lowen{S_E={{N^{4\over3}}\over{R^2}}\int{d^4x}
\partial_\mu\varphi\partial^\mu\varphi}
This gives us the expression for the four-dimensional
equilibrium distribution in the Fokker-Planck equation (50)
obtained from the full ten-dimensional dilaton action in the 
Langevin equation (18) in the s-wave approximation. Physically,
it describes the dilaton field living in the four-dimensional subspace
infinitely far from the D3-brane.

Knowing the equilibrium action $S_E$, we may now
carry out the  calculation of the drift term in (50):

\eqn\grav{\eqalign{\int{d^4p}\int{d^4q}{\delta\over{\delta\varphi(p,\rho)}}
({{\delta{S_E}}\over{\delta\varphi(q,\rho)}}P_{FP}(\rho))\cr=\int{d^4p}
\int{d^4q}\lbrace{{\delta^{2}S_E}
\over{\delta\varphi(p,\rho)\delta\varphi(q,\rho)}}
P_{FP}(\rho)+{{\delta{S_E}}\over{\delta\varphi(p,\rho)}}
{{\delta{(H_1+H_2)}}\over{\delta\varphi(q,\rho)}}P_{FP}(\rho)\rbrace\cr=
\lbrace{{N^{4\over3}}}\int{d^4}q{q^2}+
{{N^{8\over3}}\over{\rho}}\int{d^4p}\int{d^4q}
{p^2\varphi(p,\rho)q^2\varphi(q,\rho)}\cr+
{N^{8\over3}}\int{d^4p}\int{d^4q}{\partial_\rho^2\varphi(p,\rho)
q^2\varphi(q,\rho)}\rbrace{P_{FP}(\rho)}}}
Finally, substituting (68) and (72) into the right-hand  side
of the Fokker-Planck equation (50) 
we get:
\eqn\grav{\eqalign{{\hat{H}}P_{FP}(\rho)\equiv
\int{d^4p}\int{d^4q}{{\delta\over{\delta\varphi(p,\rho)}}}
\lbrack{{\delta{S_E}}\over{\delta\varphi(q,\rho)}}P_{FP}(\rho)\rbrack
\cr-\int{d^4p}\int{d^4q}\int{d\omega}{\delta\over{\delta\varphi(q,\omega)}}
G(p,\rho,q,\omega){\delta\over{\delta\varphi(p,\rho)}}P_{FP}(\rho)\cr=
-\lambda_0^6{N^{4\over3}}\partial_\rho{P_{FP}}(\rho)+\lbrace
({{N^{4\over3}}\over{R^2}}-\lambda_0^6{N^{4\over3}})\int{d^4p}p^2\cr-
(\lambda_0^6{N^{8\over3}}-{{N^{8\over3}}\over{R^2}})
\int{d^4p}\int{d^4q}{\lbrack}p^2\varphi(p,\rho)(q^2\varphi(q,\rho)+
\partial_\rho^2\varphi(q,\rho))\rbrack\rbrace{P_{FP}}(\rho)}}
Choosing $\lambda_0^6=R^{-2}$ (i.e. proportional to the curvature
of the AdS space)
 and rescaling the stochastic time 
$\rho$ parameter as $\rho=-\lambda_0^6{N^{4\over3}}s$
we see that  all the terms in the right hand side of (73)
cancel out,
except for the one proportional to the derivative of
$P_{FP}(\rho)\equiv{e^{-H_{AdS_5}(\rho)}}$:
\eqn\lowen{{\hat{H}}P_{FP}(s)={\partial\over{\partial{s}}}P_{FP}(s)}
This constitutes the proof that the $AdS_5$ gravity
indeed defines the distribution satisfying the Fokker-Planck
equation associated with the stochastic process (18) with the
non-Markovian $V_5$-noise (19).

\centerline{\bf Discussion}

Our results have the following geometrical interpretation.
 In the beginning ($\rho\rightarrow{-\infty}$)
we have an observer living in asymptotically flat
four-dimensional space-time. At certain moment the D3-brane,
or the $V_5$ vertex, located infinitely far from the observer, is introduced.
 The gravitational force of the brane causes  the drift of the
four-dimensional space-time slice, in which the observer lives,
towards the D3-brane location. From the point of view of the observer,
however, the process of the four dimensional
space-time slice deformation is perceived
as the non-Markovian stochastic process with the white noise propagating
in the additional fictitious dimension. However, what perceived
by the observer as a stochastic drift in the fictitious time dimension
is in fact the gravitational drift in the $real$ dimension, that is,
 the coordinate transverse to the $D3$-brane worldvolume.
The equilibrium  state in the limit of  
$\rho\rightarrow\infty$
corresponds to reaching the D3-brane horizon by the observer.
In this limit the $V_5$ noise propagator (51) becomes
\eqn\lowen{lim_{\rho\rightarrow\infty}G(p,\rho,q,\omega)=
\lambda_0^4+\lambda_0^6}
and in this limit the kinetic term of the Fokker-Planck equation (50)
vanishes. The only contribution comes then from the drift term and
the Fokker-Planck equation (50) is reduced to
${p^2}P_{FP}(\varphi)\sim{0}$ or
\eqn\lowen{{{\partial^2}\over{\partial{X_\mu}\partial{X^\mu}}}P_{FP}\sim{0}}
in the position space. The structure of the last equation is similar
to the regularized large N loop equation 
\eqn\grav{\eqalign{{\hat{L}}W(C)=0}}
with
\eqn\lowen{{\hat{L}}\equiv{lim_{\epsilon\rightarrow{0}}}
{\int_{-\epsilon}^{\epsilon}}
{d\alpha}
{{\delta^2}\over{\delta{X_\mu}(s-\alpha)\delta{X^\mu}(s+\alpha)}}}
Therefore in the limit of infinite stochastic time the 
Fokker-Planck distribution $e^{-H_{AdS_5}(\rho)}$ solving the equation (50)
 is  reduced to
the Wilson loop of the large N gauge theory living in 
the D3-brane worldvolume. From  point of view of the Langevin equation
(18) reaching the equilibrium limit corresponds to restoring  the
worldsheet conformal invariance (or cutoff independence).
Our final remark regards higher order $\lambda_0$ corrections in
the brane-like sigma-model (9). 
As has been said before, this four-dimensional picture is only appropriate
in the dilaton s-wave approximation.  Higher order $\lambda_0$
corrections correspond to higher angular momentum modes
of the dilaton on $S^5$, therefore the initial equilibrium distribution
 in (18) must be taken ten-dimensional.
Just as described above, the  stochastic process is non-Markovian
and the equilibrium limit reached at infinite stochastic time is
different from the initial one and it corresponds to the
$AdS_5\times{S^5}$ gravity. Summing up to all orders of $\lambda_0$
one should obtain string theory compactified on the $AdS_5\times{S^5}$
space. Thus flat ten-dimensional string theory and the 
one in the $AdS_5{\times}{S^5}$ space-time appear as two
$different$ thermodynamical limits of the  stochastic
process caused by the non-Markovian $V_5$-noise. This picture offers
the dynamical explanation of how
the $AdS_5\times{S^5}$ space-time geometry emerges in the brane-like
sigma-model (5) due to the presence of the 5-form $V_5$ vertex operator.
Another question of interest is the loop equation which open string
theory in the $AdS_5\times{S^5}$ background should satisfy ~\refs{\ampr}.
Though Wilson loops have been studied extensively in the context of AdS/CFT
correspondence (for review see ~\refs{\sonn}), it is not clear
that the functional of the $AdS_5$ supergravity satisfies the loop equation.
So far the zig-zag invariance of the $AdS_5$ string has
 been proved in the WKB limit and
for the special class of contours only ~\refs{\ampr}. On the other hand,
it has been argued ~\refs{\periwal}
that the loop equation may be understood as an equilibrium condition for
the Fokker-Planck equation with the white noise.
It seems hard, however, to  draw any straightforward connection between
Fokker-Planck and Schwinger-Dyson equations and, in particular,
to interpret the loop equation as an infinite time limit of
the $<H_{FP}(\tau)W(C)>=0$ relation discussed in ~\refs{\periwal,
\hirano, \van}.
The reason is that the loop equation is derived in the regularized theory
while no corresponding regularization is performed in the
Fokker-Planck counterpart. Therefore in order to
 to make sense of possible connections
between Wheeler-de Witt and Schwinger-Dyson equations one has to
obtain more information about string and gauge
degrees of freedom and their interrelation.
 In particular, it is necessary to point out how string
and gauge theory cutoffs correspond to each other.
We hope that our approach, attempting to explore the holographic effects
from  brane-like sigma-model point of view, will be helpful in trying to
find answers to these questions.

\centerline{\bf Acknowledgements}
It is a  pleasure to thank the Physics Department of Tel Aviv University
and particularly J.Sonnenschein for their  gracious hospitality.
The author also gratefully acknowledges the very kind hospitality of 
the Erwin  Schroedinger  International Institute for Mathematical Physics
as well as the organizers of the workshop ``Dualities in String Theory''
at the ESI in Vienna. In particular, it is a pleasure to thank
H. Grosse and S. Theisen. I also would like to thank
H.Huffel, B.Kol, M. Sheikh-Jabbari,
J.Sonnenschein and S.Theisen for useful discussions.
This work is partially supported by the EC contract 
no. ERBFMRX-CT 96-0090.

\listrefs
\end